\pdfoutput=1

\documentclass[11pt,twoside,a4paper,cmspaper,final,collab]{cms-tdr}
\def\svnVersion{266931}\def\cmsCernNo{2013-037}\def\cmsCernDate{\today}\def\cmsMessage{Published in Physical Review D as \href{http://dx.doi.org/10.1103/PhysRevD.85.112007}{\doi{10.1103/PhysRevD.85.112007.}}}

\begin{document}\cmsNoteHeader{TOP-11-006}

\hyphenation{had-ron-i-za-tion}
\hyphenation{cal-or-i-me-ter}
\hyphenation{de-vices}
\RCS$Revision: 120025 $
\RCS$HeadURL: svn+ssh://svn.cern.ch/reps/tdr2/papers/TOP-11-006/trunk/TOP-11-006.tex $
\RCS$Id: TOP-11-006.tex 120025 2012-05-05 17:08:13Z michgall $
\newcommand\pfJET{\textsc{pfJET}\xspace}
\newcommand\pfMET{\textsc{pfMET}\xspace}
\newcommand\htmiss{\ensuremath{H_\mathrm{T}^\text{miss}\xspace}}
\newcommand\tauh{\ensuremath{\Pgt_\mathrm{h}}}
\ifthenelse{\boolean{cms@external}}{\providecommand{\cmsLeft}{top}}{\providecommand{\cmsLeft}{left}}
\ifthenelse{\boolean{cms@external}}{\providecommand{\cmsRight}{bottom}}{\providecommand{\cmsRight}{right}}
\cmsNoteHeader{TOP-11-006}
\title{\texorpdfstring{Measurement of the \ttbar production cross section
in pp collisions at $\sqrt{s}$=7\TeV
in dilepton final states containing a $\Pgt$}{Measurement of the top quark pair production cross section in pp collisions at sqrt(s) = 7 TeV
in dilepton final states containing a tau}
}

\date{\today}

\abstract{
The top quark pair production cross section is measured in dilepton events with one
electron or muon, and one hadronically decaying $\Pgt$ lepton from the decay
$\ttbar\rightarrow (\ell\Pgn_\ell) (\tauh \Pgngt) {\bbbar}$, ($\ell=\Pe, \Pgm$).
The data sample corresponds to an integrated luminosity of 2.0\fbinv for the electron channel and 2.2\fbinv for the muon channel,
collected by the CMS detector at the LHC.
This is the first measurement of the \ttbar cross section explicitly including
$\Pgt$ leptons in proton-proton collisions at $\sqrt{s}=7$\TeV.
The measured value $\sigma_{\ttbar} = 143  \pm 14 (\text{stat.}) \pm 22 (\text{syst.}) \pm 3 (\text{lumi.})\unit{pb}$
is consistent with the standard model predictions.
}

\hypersetup{
pdfauthor={CMS Collaboration},
pdftitle={Measurement of the top quark pair production cross section in pp collisions at sqrt(s) = 7 TeV
in dilepton final states containing a tau},
pdfsubject={CMS},
pdfkeywords={CMS, LHC, physics, top quark, taus}}

\maketitle

\newcommand{\mymet}{\makebox[2.4ex]{\ensuremath{\not\!\! E_{\mathrm{T}}}}}
\newcommand{\myttbar}{\ensuremath{{t\overline{t}}}\xspace}

\section{Introduction}

Top quarks at the Large Hadron Collider (LHC) are mostly produced in pairs with the subsequent decay $\ttbar\rightarrow \PWp\cPqb\PWm\cPaqb$.
The decay modes of the two W bosons determine the observed event signature.
The dilepton decay channel denotes the case where both W bosons from the decaying top quark pair decay leptonically.
In this Letter, top quark decays in the ``tau dilepton" channel are studied, where one \PW\ boson decays into $\Pe\Pgn$ or $\Pgm\Pgn$
and the other into the hadronically decaying $\Pgt$ lepton and $\Pgn$,
in the final state $\ttbar\rightarrow (\ell\Pgn_\ell) (\tauh \Pgngt) \bbbar$, where $\ell= \Pe, \Pgm$.
The expected fraction of events
in the dilepton channel with at least one $\Pgt$ lepton in the final state is approximately 6\% (5/81) of all \ttbar decays,
i.e. higher than the fraction of the light dilepton channels (\Pe\Pe, $\Pgm\Pgm$, $\Pe\Pgm$) which is equal to 4/81 of all \ttbar decays.
The tau dilepton channel is of particular interest because the existence of a charged Higgs boson~\cite{Gunion:1989we,Djouadi:2005gj} with a mass smaller
than the top quark mass
could give rise to anomalous $\Pgt$ lepton production, which could be directly observable in this decay channel.
Furthermore, in the final state studied, the $\cPqt\rightarrow (\Pgt\Pgngt) \cPqb$ decay exclusively involves third generation leptons and quarks.
Understanding the $\Pgt$ yield in top quark decays is important
to increase the acceptance for \ttbar events and to search for new physics processes.

This is the first measurement of the \ttbar production cross section at the LHC that explicitly includes $\Pgt$ leptons,
improving over the results obtained at the Tevatron which are limited by the small number of candidate events found~\cite{c:taudil_cdf1,c:taudil_cdf2,c:d0taudil}.
Experimentally, the $\Pgt$ lepton is identified by its decay products, either hadrons ($\tauh$) or leptons ($\Pgt_\ell$), with the corresponding branching fractions
$Br(\tauh \rightarrow \text{hadrons} + \Pgngt) \simeq 65\%$ and
$Br(\Pgt_\ell \rightarrow \ell~\Pgn_{\ell}\Pgngt, \ell = \mathrm{e}, \Pgm) \simeq 35\%$.
In the first case, a narrow jet with a distinct signature is produced; in the case of leptonic decays, the distinction
from prompt electron or muon production is experimentally difficult, consequently only hadronic $\Pgt$ decays are studied here.
The cross section is measured by counting the number of $\Pe\tauh+X$ and $\Pgm\tauh+X$ events consistent with originating
from \ttbar, subtracting the contributions from other processes, and correcting for the efficiency of the event selection.
The measurement is based on data collected by the Compact Muon Solenoid (CMS)
experiment in 2011.  The integrated luminosity of the data samples
are 1.99\fbinv and 2.22\fbinv for the $\Pe\tauh$ and $\Pgm\tauh$ final states, respectively.

The CMS detector is briefly summarized in Section~\ref{sec:detector},
details of the simulated samples are given in Section~\ref{sec:simulation},
a brief description of the event reconstruction and event selection
is provided in Section~\ref{sec:eventsel},
followed by the description of the background determination and systematic uncertainties in
Sections~\ref{sec:background} and~\ref{sec:systematics}, respectively.
The measurement of the cross section is discussed in Section~\ref{sec:xsec}, and the
results are summarized in Section~\ref{sec:summ}.

\section{The CMS detector}
\label{sec:detector}

The central feature of the CMS apparatus is a superconducting solenoid,
13\unit{m} in length and 6\unit{m} in diameter, which provides an axial magnetic
field of 3.8\unit{T}.  Inside the solenoid, various
particle detection systems are employed.  Charged particle trajectories are
measured by the silicon pixel and strip tracker, covering $0 < \varphi <
2\pi$ in azimuth and $|\eta |<$~2.5, where the pseudorapidity $\eta$ is
defined as $\eta =-\ln[\tan ({\theta/2}) ]$, with $\theta$ being the
polar angle of the trajectory of the particle with respect to the
counterclockwise beam direction.
A crystal electromagnetic calorimeter and a brass/scintillator hadron calorimeter surround the tracking volume;
in this analysis the calorimetry provides high-resolution energy and direction measurements of electrons and hadronic jets.
Muon detection systems are located outside of the solenoid and embedded in the steel return yoke.
The detector is nearly hermetic, allowing for energy balance
measurements in the plane transverse to the beam directions.
A two-level trigger system selects the most interesting proton-proton collision events for use in physics analysis.
A more detailed description of the CMS detector can be found elsewhere~\cite{JINST}.

\section{Event simulation}
\label{sec:simulation}

The analysis makes use of simulated samples of \ttbar events as well as other processes that result in \Pgt{}s in the final state.
These samples are used to design the event selection, to calculate the acceptance to \ttbar events, and to estimate some of the backgrounds in the analysis.

Signal \ttbar events are simulated with the \MADGRAPH event generator (v.~5.1.1.0)~\cite{madgraph}
with matrix elements corresponding to up to three additional partons, for a top quark mass of 172.5~\GeVcc.
The number of expected \ttbar events is estimated with the approximate next-to-next-to-leading order (NNLO) expected standard model (SM) cross section value
of $165^{+4}_{-9} (\text{scale}) ^{+7}_{-7} (\mathrm{PDF})$\unit{pb}~\cite{ttbarxsec1,ttbarxsec2},
where the first uncertainty is due to renormalization and factorization scales, and the second is due to the parton distribution function (PDF) uncertainty.
This cross section is used for illustrative purposes to normalize the \ttbar $\Pe\tauh$ and $\Pgm\tauh$ expectations discussed in Section~\ref{sec:eventsel}.
The generated events are subsequently processed with \PYTHIA (v.~6.422)~\cite{pythia}
to provide the showering of the partons, and to perform the matching of the soft radiation
with the contributions from direct emissions accounted for in the matrix-element calculations.
The Z2 tune~\cite{c:z2} is used with the CTEQ6L PDFs~\cite{pdfset}.
The $\Pgt$ decays are simulated with \TAUOLA (v.~27.121.5)~\cite{tauola}
which correctly accounts for the $\Pgt$
lepton polarization in describing the decay kinematics.
The CMS detector response is simulated with \GEANTfour (v.~9.3 Rev01)~\cite{geant}.

The background samples used in the measurement of the cross section are simulated with \MADGRAPH and \PYTHIA.
The W+jet samples include only the leptonic decays of the \PW\ boson,
and are normalized to the inclusive next-to-next-leading-order (NNLO) cross section of  $31.3\pm 1.6$~nb,
calculated with the \textsc{fewz} (Fully Exclusive \PW\ and \cPZ\ boson) production program~\cite{fewz}.
Drell--Yan (DY) pair production of charged leptons in the final state
is generated with \MADGRAPH for dilepton invariant masses above 50\GeVcc,
and  is normalized to a cross section of $3.04\pm0.13$\unit{nb},
computed with {\sc fewz}.
The DY events with masses between 10 and 50\GeVcc are generated with \MADGRAPH
with a cross section (with a k-factor of 1.33 to correct for NLO) of 12.4~nb.

The electroweak production of single top quarks is considered as a background process, and is simulated with \POWHEG~\cite{powheg}.
The $t$-channel single top quark NLO cross section is $\sigma_{t-\mathrm{ch.}} = 64.6^{+3.4}_{-3.2}$\unit{pb}
from \textsc{mcfm}~\cite{mcfm,mcfm2,mcfm3,mcfm4}.
The single top quark associated production (\cPqt\PW) cross section amounts to
$\sigma_{\cPqt\PW}=15.7\pm1.2$\unit{pb}~\cite{Kidonakis:2010tW}.
The $s$-channel single top quark next-to-next-leading-log (NNLL) cross section is determined as
$\sigma_{s-\mathrm{ch.}} = 4.6\pm0.06$\unit{pb}~\cite{NNLLtop}.
Finally, the production of \PW\PW, \PW\cPZ, and \cPZ\cPZ\ pairs, with
inclusive cross sections
of $43.0\pm1.5$\unit{pb}, $18.8\pm 0.7$\unit{pb}, and $7.4\pm 0.2$\unit{pb}, respectively
(all calculated at the NLO with \textsc{mcfm}), are simulated with \PYTHIA.

\section{Event selection}
\label{sec:eventsel}

The signal topology is defined by the presence of two $\cPqb$ jets from the top quark decays, one \PW\ boson decaying
leptonically into $\Pe\Pgn$ or $\Pgm\Pgn$, and a second \PW\ boson decaying into $\Pgt\nu$.
In the event, all objects are reconstructed with a particle-flow (PF) algorithm~\cite{c:pft-09-001}.
The PF algorithm combines the information from all sub-detectors to identify and
reconstruct all types of particles produced in the collision, namely charged hadrons, photons, neutral
hadrons, muons, and electrons. The resulting list of particles is used to construct a
variety of higher-level objects and observables such as jets, missing transverse energy (\MET),
leptons (including $\Pgt$s), photons, \cPqb-tagging discriminators, and isolation variables.
The missing transverse energy \MET is computed as the absolute value of the vectorial sum
of the transverse momenta of all reconstructed particles
in the event.

Electron or muon candidates are required to be isolated relative to other activity in the event.
The relative isolation is based on PF objects and defined as
$ I_\text{rel} = (E_\text{ch}+ E_\mathrm{nh}+ E_\mathrm{ph})/\pt \cdot c$, where
$E_\text{ch}$ is the transverse energy deposited by charged hadrons in a cone of radius $\Delta R=0.3$ around the electron or muon track,
$E_\mathrm{nh}$ and $E_\mathrm{ph}$ are the respective transverse energies of the neutral hadrons and photons, 
and \PT is the electron or muon transverse momentum.
The electron (muon) candidate is considered to be non-isolated and is rejected if $I_\text{rel} > 0.1$ ($>0.2$).
Jets are reconstructed with the anti-$k_\mathrm{T}$~\cite{antikt,fastjet} jet algorithm with a distance parameter $R=0.5$.

Hadronic $\Pgt$ decays are reconstructed with the Hadron Plus Strips (HPS) algorithm~\cite{c:tau-11-001}.
The identification process starts with the clustering of all PF particles into jets with the
anti-{$k_\mathrm{T}$} algorithm with a distance parameter $R = 0.5$.
For each jet, a charged hadron is combined with other nearby charged hadrons or photons to identify the decay modes.
The identification of $\Pgpz$ mesons is enhanced by clustering electrons and photons in
``strips" along the bending plane to take into account possible broadening of calorimeter signatures by early showering photons.
Then, strips and charged hadrons are combined to reconstruct the following combinations: single
hadron, hadron plus a strip, hadron plus two strips and three hadrons.
To reduce the contamination from quark and gluon jets, the $\tauh$ candidate isolation is calculated
in a cone of $\Delta R = 0.5$ around the reconstructed $\Pgt$-momentum direction.
It is required that there be no charged hadrons with $\PT > 1.0$\GeVc and
no photons with $\ET > 1.5$\GeV in the isolation cone, other than the $\Pgt$ decay particles.
Additional requirements are applied to discriminate genuine $\Pgt$ leptons from prompt electrons and muons.
The $\Pgt$ charge is taken as the sum of the charges of the charged hadrons (prongs) in the signal cone;
its uncertainty is less than 1\% and it is estimated from same sign  $Z\rightarrow \tau\tau\rightarrow\mu\tauh$ events~\cite{c:sus-11-010}.
The $\Pgt$ reconstruction efficiency of this algorithm is estimated to be approximately 37\%
(i.e. ``medium" working point in Ref.~\cite{c:tau-11-001}) for $\pt^{\tauh}>20\GeVc$, and
it is measured in a sample enriched in $Z\rightarrow\Pgt\Pgt\rightarrow\Pgm\tauh$ events
with a ``tag-and-probe" technique~\cite{ewk-10-005}.
The ``medium" working point corresponds to a probability of approximately 0.5\% for generic hadronic jets to be misidentified as $\tauh$.

For the $\Pe\Pgt_{h}$ final state, events are triggered by the combined electron
plus two jets
plus $\htmiss$
trigger ($\Pe+\text{dijet}+\htmiss$),
where $\htmiss$ is the absolute value of the vectorial sum of all jet momenta in the plane transverse to the beams.
The thresholds for the electron and for $\htmiss$ are respectively $\pt>$17--27\GeVc and $\htmiss>$15--20\GeV
depending on the data-taking period,
and the \pt thresholds for the two jets are 30\GeVc and 25\GeVc.
The trigger efficiency  is estimated from a suite of triggers with lower thresholds
assuming the factorization $\epsilon_\text{trig} = \epsilon_\text{e} \times \epsilon_\text{jets} \times \epsilon_\mathrm{MHT}$,
where $\epsilon_\Pe$ is the electron efficiency, $\epsilon_\text{jets}$ is the efficiency for selecting two jets,
and $\epsilon_\mathrm{MHT}$ is the efficiency for $\htmiss$.
The data-to-simulation scale factor for the electron trigger efficiency is $0.99\pm 0.01$.
The efficiencies $\epsilon_\mathrm{MHT} = 1.00^{+0.00}_{-0.01}$ and $\epsilon_\text{jets}$, which is parameterized as a function of jet \pt, are estimated from data.
In the $\Pgm\tauh$ final state,
data are collected with a trigger requiring at least one isolated muon with threshold of $\pt>17 (24)$\GeVc, for the earlier (later) part of the data sample;
the data-to-simulation scale factor for the trigger efficiency is $0.99\pm 0.01$.

Events are selected by requiring one isolated electron (muon)
with transverse momentum $\pt>35 (30)$\GeVc and $|\eta |<2.5 (2.1)$, at least two jets with
$\pt>35 (30)$\GeVc and $|\eta |<2.4$, missing transverse energy \MET$>45(40)$\GeV and one
hadronically decaying $\Pgt$ lepton ($\Pgt$ jet) with $\pt> 20$\GeVc and $|\eta |<2.4$.
Electrons or muons are required to be separated from any jet in the
($\eta, \varphi$) plane by a distance $\Delta R > 0.3$. Events with any additional
loosely isolated ($I_\text{rel} < 0.2$) electron (muon) of $\pt>15~(10)$\GeVc are rejected.

The $\Pgt$ jet and the lepton are required to have electric charges of opposite sign (OS).
At least one of the jets is required to be identified as originating from \cPqb\ quark hadronization (\cPqb\ tagged).
The \cPqb-tagging algorithm used (``TCHEL" in Ref.~\cite{c:btv-11-001}) is based on sorting tracks according to
their impact parameter significance ($S_\mathrm{IP}$); the $S_\mathrm{IP}$ value of the second track is used as the discriminator.
The \cPqb-tagging efficiency of this algorithm is $76\pm 1$\%, measured in a sample of events enriched with jets from semileptonic \cPqb-hadron decays.
The misidentification rate of light-flavor jets is obtained from inclusive jet studies and is measured to be $13\pm 3$\% for jets
in the \pt range relevant to this analysis.
After the final event selection, a fraction of approximately 12\% of the generated \ttbar tau dilepton events
within the geometric and kinematic fiducial region are selected.

\begin{figure}[htp]
\begin{center}
\includegraphics[width=0.45\textwidth]{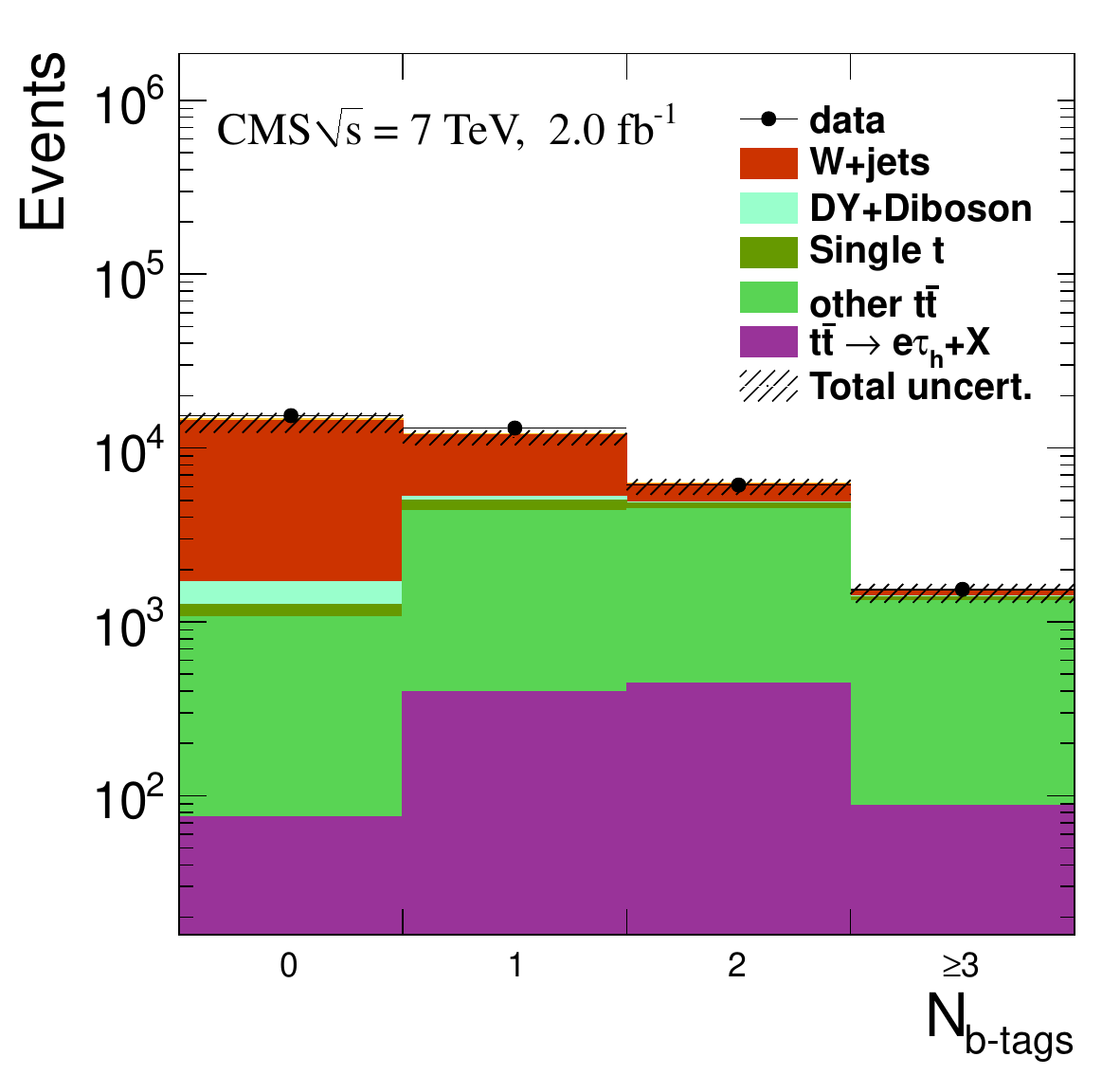}\hfill
\includegraphics[width=0.45\textwidth]{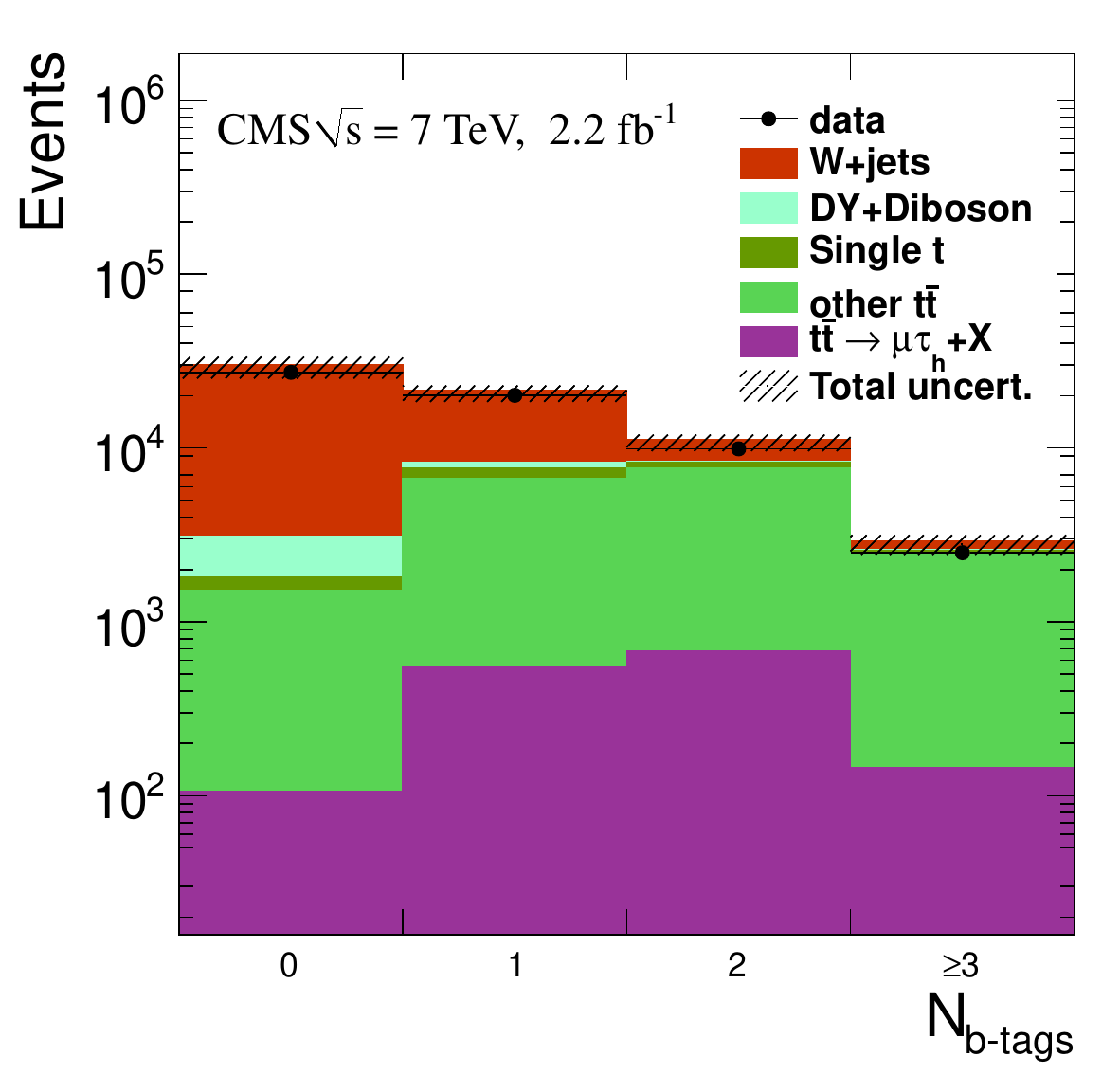}
\caption{
The \cPqb-tagged jet multiplicity for pre-selected events with
one electron (\cmsLeft) or muon (\cmsRight).
Distributions obtained from data (points) are compared with simulation.
The simulated contributions are normalized to the SM predicted values. The hatched area shows the total systematic uncertainty.
}
\label{fig:jetmultiplicity}
\end{center}
\end{figure}

\begin{figure}[htp]
\begin{center}
\includegraphics[width=0.45\textwidth]{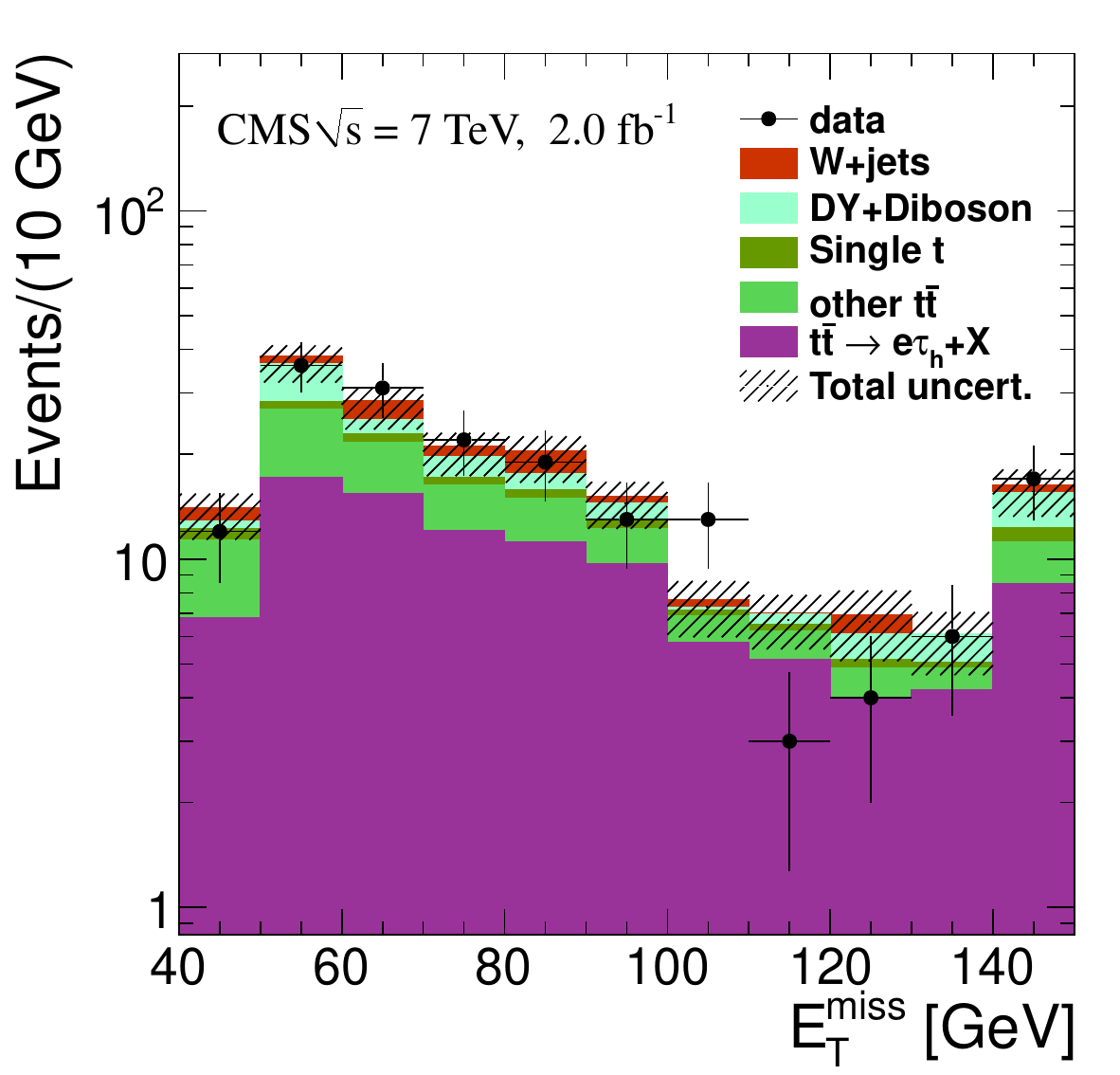} \hfill
\includegraphics[width=0.45\textwidth]{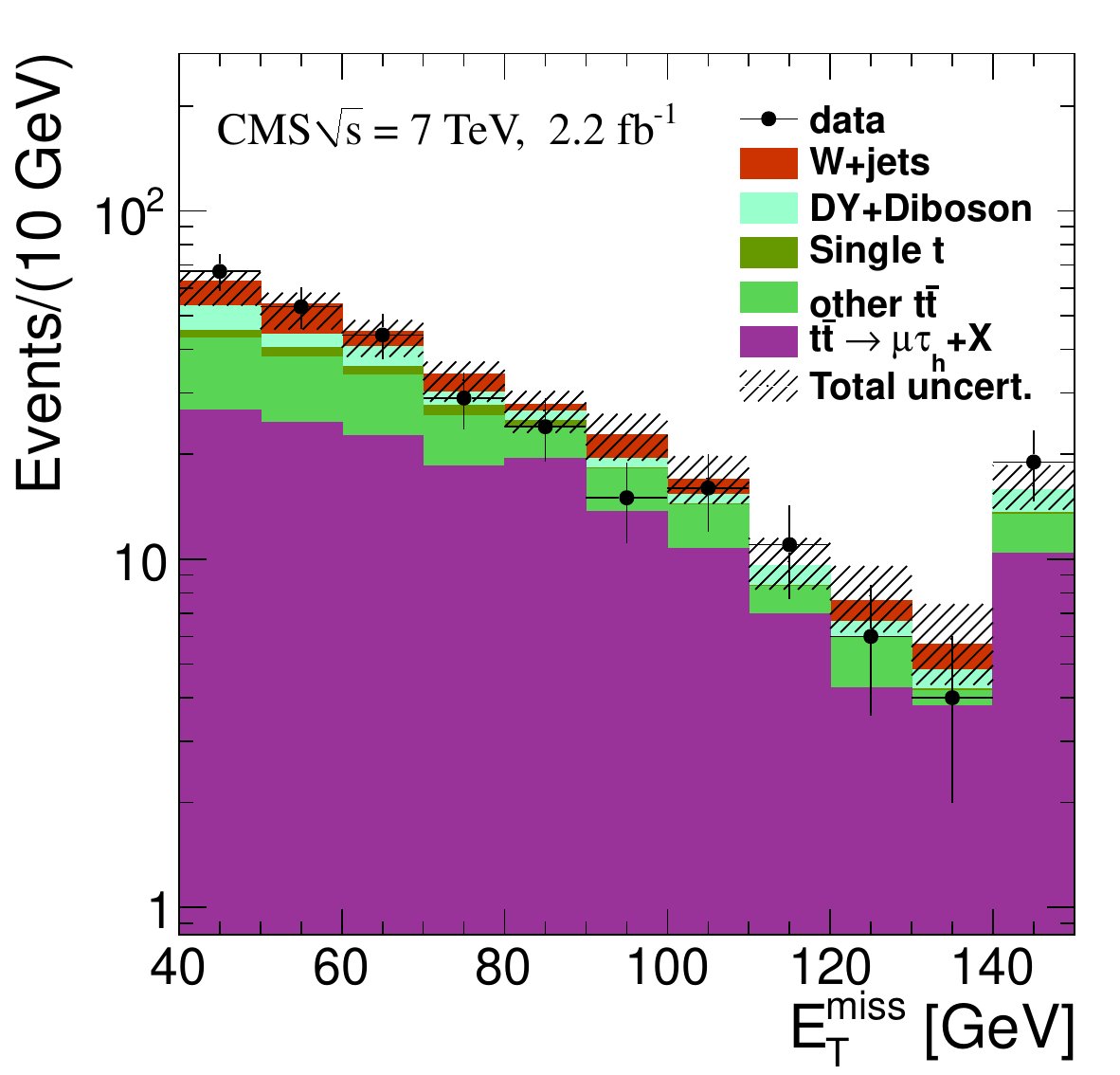}
\caption{
\MET distribution after the full event selection for the $\Pe\tauh$ (\cmsLeft)
and $\Pgm\tauh$ (\cmsRight) final states.
Distributions obtained from data (points) are compared with simulation.
The last bin includes the overflow.
The simulated contributions are normalized to the SM predicted values.
The hatched area shows the total systematic uncertainty.
}
\label{fig:metdistribution}
\end{center}
\end{figure}

\begin{figure}[htp]
\begin{center}
\includegraphics[width=0.49\textwidth]{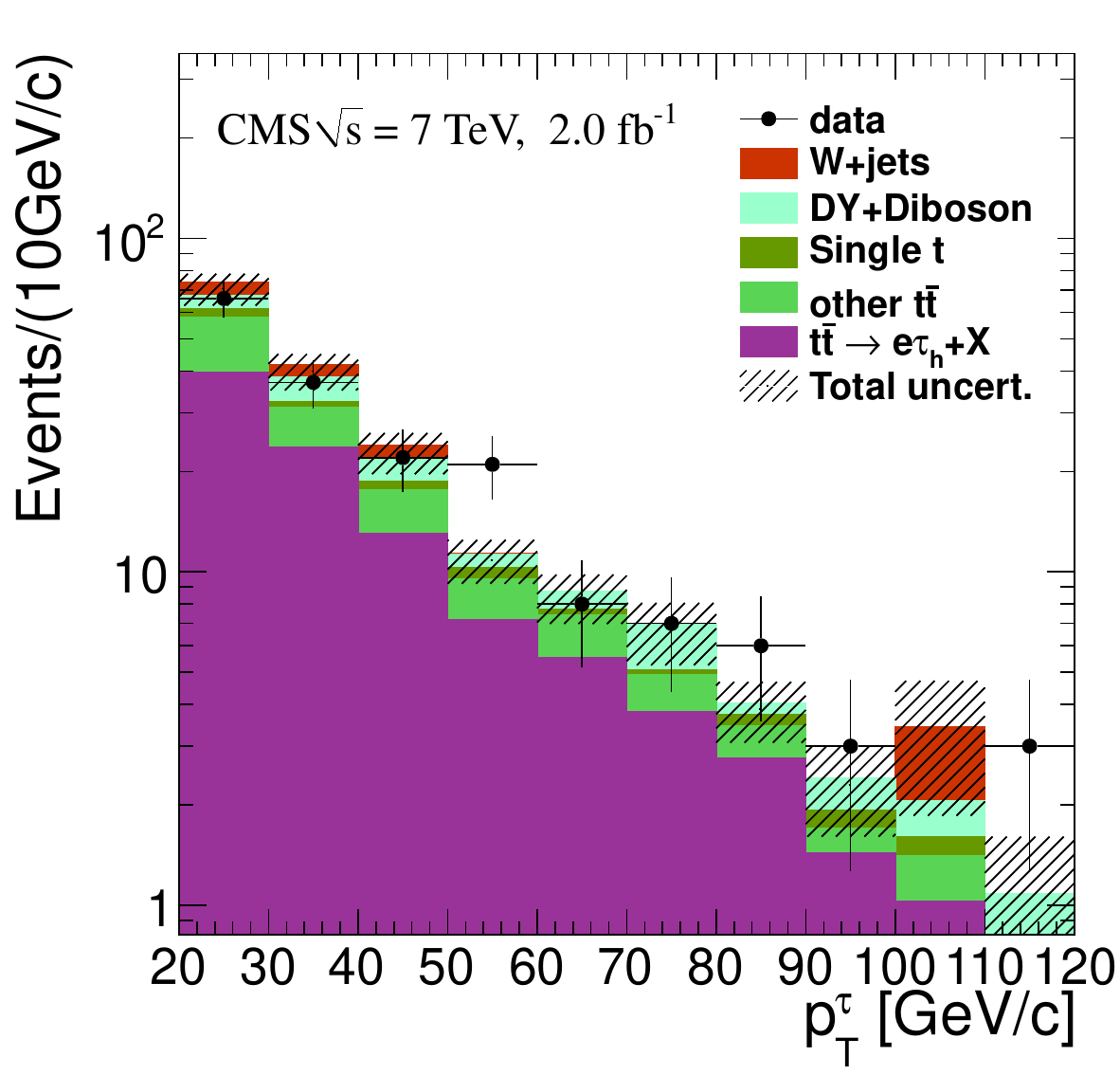}
\includegraphics[width=0.49\textwidth]{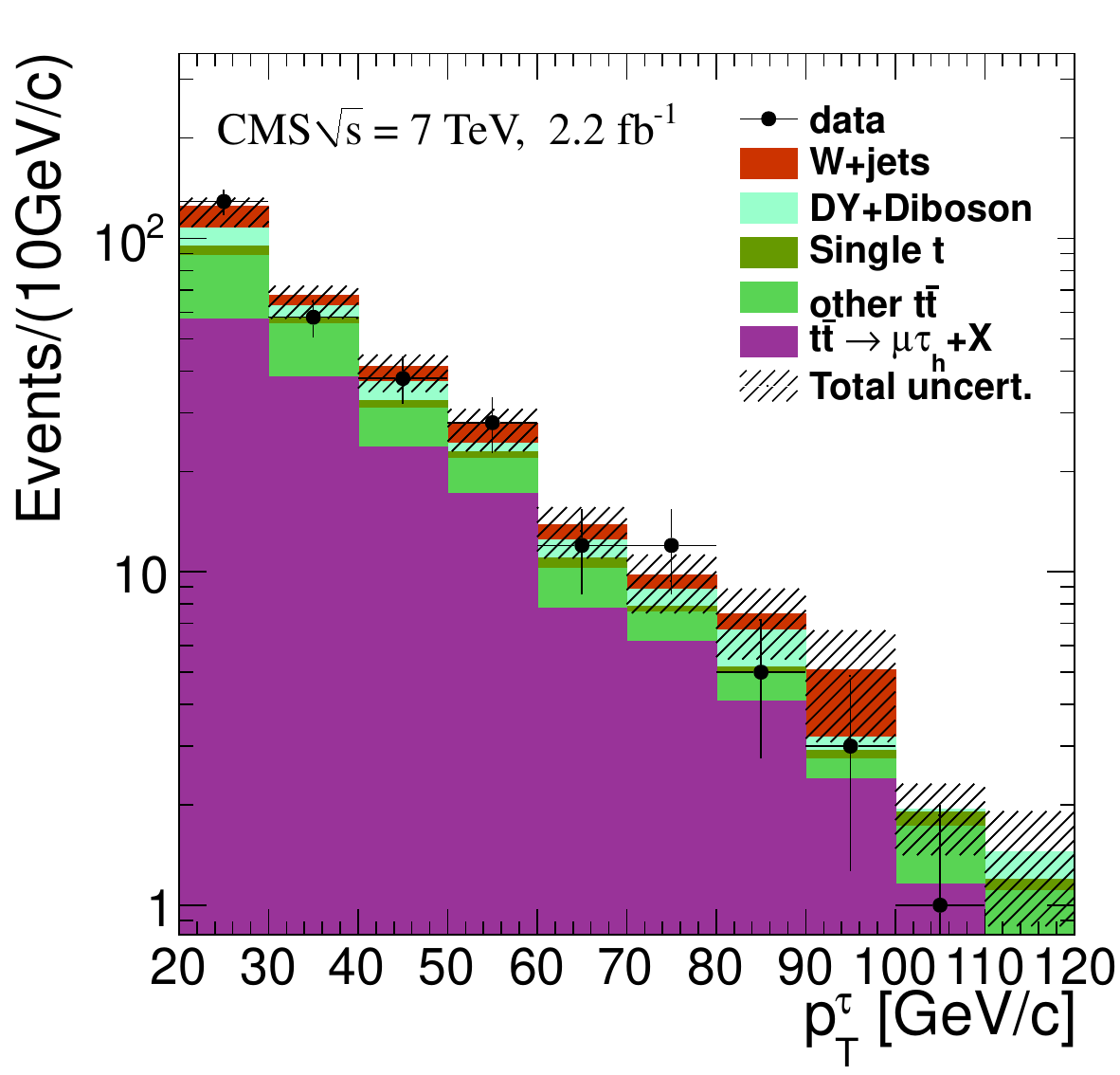}
\caption{
The $\Pgt$ \pt distribution after the full event selection for the $\Pe\tauh$ (\cmsLeft)
and $\Pgm\tauh$ (\cmsRight) final states.
Distributions obtained from data (points) are compared with simulation.
The simulated contributions are normalized to the SM predicted values.
The hatched area shows the total systematic uncertainty.
}
\label{fig:finaldistributions_tau}
\end{center}
\end{figure}

The \cPqb-tagged jet multiplicity for the $\Pe\tauh$ and
$\Pgm\tauh$ final states is shown in Fig.~\ref{fig:jetmultiplicity}
for the events in the pre-selected sample,
i.e. one isolated electron (muon), missing transverse energy above 45~(40)\GeV, and at least three jets, two jets with $\pt>35 (30)$\GeVc and one jet with $\pt> 20$\GeVc.
The observed numbers of events are consistent with the expected numbers of signal and background events obtained from the simulation.
The distributions of the \MET and of the transverse momentum of the $\Pgt$ lepton
after the final event selection
are shown in Fig.~\ref{fig:metdistribution} and in Fig.~\ref{fig:finaldistributions_tau}, respectively, for both the $\Pe\tauh$ and $\Pgm\tauh$ final states.
The distributions show good agreement between the observed numbers of events and the expected numbers of signal and background events obtained from the simulation.
The \MET distribution for the $\Pe\tauh$ final state has a deficit of events in the first bin due to the higher \MET threshold, when compared to the $\Pgm\tauh$ final state.

\begin{figure}[htp]
\begin{center}
\includegraphics[width=0.49\textwidth]{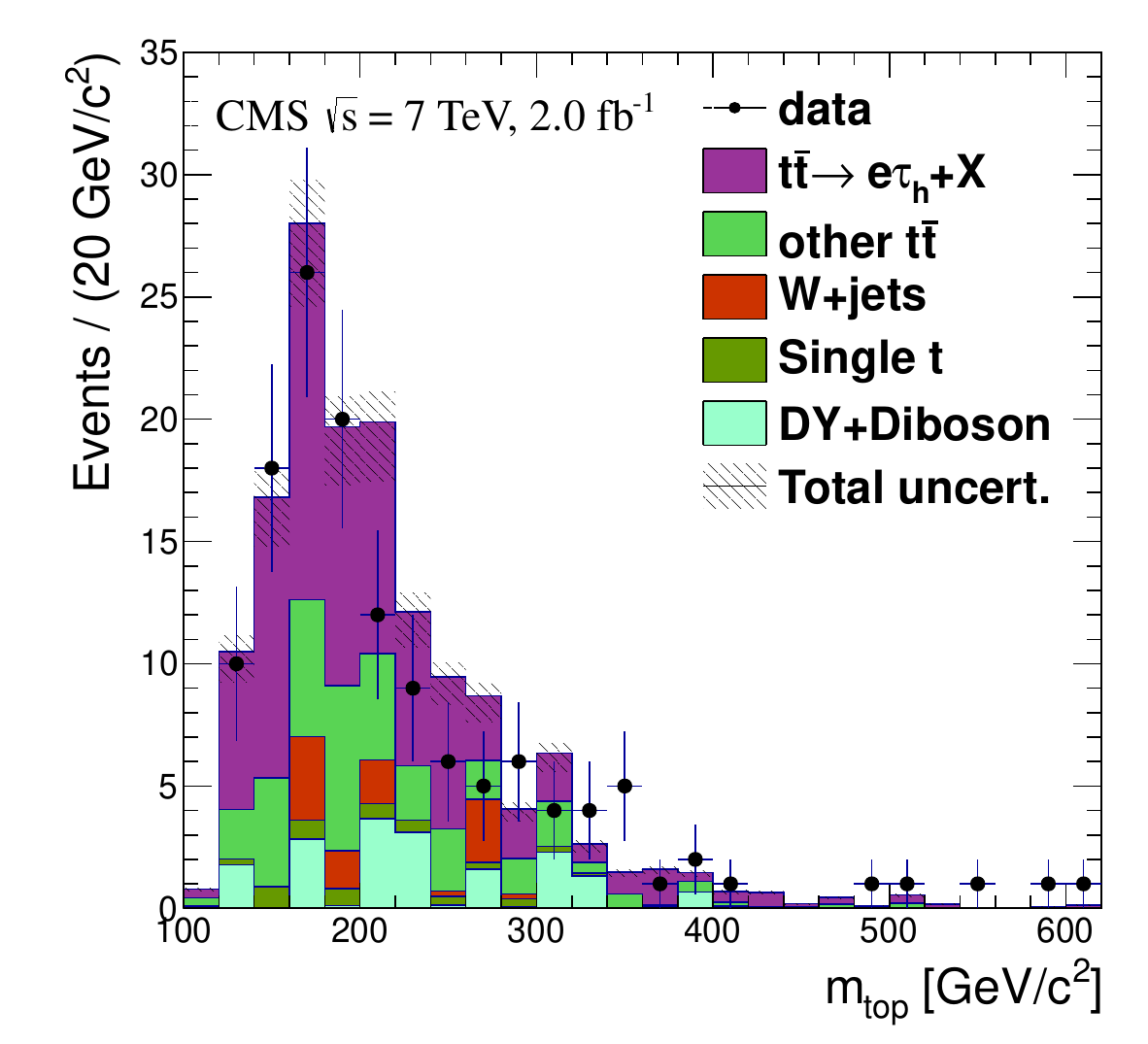} \hfill
\includegraphics[width=0.49\textwidth]{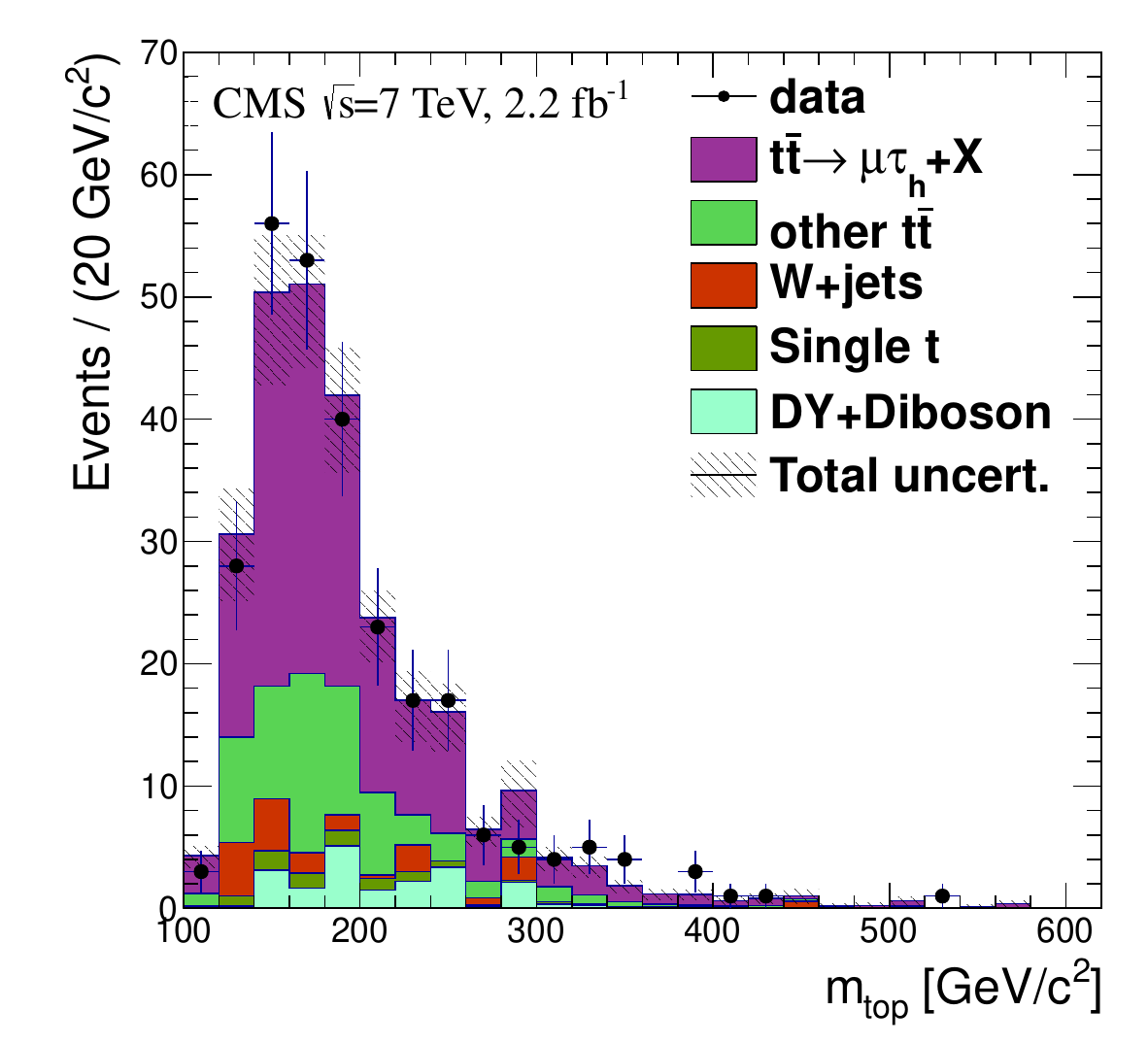}
\caption{
Reconstructed top quark mass $m_\text{top}$ distribution for the $\Pgt$ dilepton candidate events after the full event selection,
in the $\Pe\tauh$ (\cmsLeft) and $\Pgm\tauh$ (\cmsRight) final states.
Distributions obtained from data (points) are compared with simulation.
The hatched area shows the total systematic uncertainty.
}
\label{fig:taudileptonmtop}
\end{center}
\end{figure}

The top quark mass is reconstructed with the KINb~\cite{c:top-11-002}
algorithm (Fig.~\ref{fig:taudileptonmtop}), treating the additional neutrino
in the $\Pgt$ decay as a contribution to the \MET.
Numerical solutions for the kinematic reconstruction of \ttbar decays
with two charged leptons in the final state are found for each event.
The jet transverse momentum, the \MET direction, and the longitudinal momentum of the \ttbar system
are varied independently within their measured resolutions
to scan the kinematic phase space compatible with the \ttbar system.
Solutions with the lowest invariant mass of the \ttbar system are accepted if the difference between the two top quark masses is less than 3~\GeVcc.
The reconstructed top quark mass in Fig.~\ref{fig:taudileptonmtop} shows that the kinematic properties of the selected events
are statistically compatible with predictions based on a top quark mass of 172.5\GeVcc,
indicating the consistency of the selected sample in data
with the sum of top quark pair production plus the background.

\section{Background estimate}
\label{sec:background}

The background comes from two categories of events, the ``misreconstructed $\Pgt$" background ($N^\text{misid}$)
which is estimated from data, and the
``other" background ($N^\text{other}$) which is estimated from simulation.

The main background (misreconstructed $\Pgt$) comes from events with one lepton (electron or muon), \MET requirement and three or more jets,
where one jet is misidentified as a $\Pgt$ jet. 
The dominant contribution to this background is from events where one W boson is produced in association with jets,
and from $\ttbar \rightarrow {\PWp\cPqb \PWm\cPaqb}\rightarrow \ell\Pgn {\cPqb\ ~\Pq\cPaq^\prime\cPaqb}$ events.
In order to estimate this background from data, the probability $w(\text{jet} \rightarrow \tauh)$ that a jet is misidentified as 
a $\Pgt$ jet as a function of the jet \pt, $\eta$, and jet width ($R_\text{jet}$)
is determined, then applied to every jet in the pre-selected sample with one \cPqb-tagged jet.
The quantity $R_\text{jet}$ is defined as $\sqrt{\sigma_{\eta\eta}^2 + \sigma_{\phi\phi}^2 }$, where
$\sigma_{\eta\eta}$ ($\sigma_{\phi\phi}$) expresses the extent in $\eta$ ($\phi$) of the jet cluster.
Thus the expected number of background is obtained as:

\begin{equation} \label{eq:fakes}
N^\text{misid} = \sum^{N}_{i}~ \sum ^{n}_{j} w_{i}^{j} (\text{jet} \rightarrow \Pgt) - N^\text{other},
\end{equation}

where $j$ is the jet index of the event $i$.
The quantity $N^\text{other}$ is the small (${\simeq}18 \%$) contamination of other
contributions to the misidentified $\Pgt$ background, which is estimated from simulation.
This is mostly due to the presence of genuine $\Pgt$ jets in the $\PW+\ge 3$~jet sample.
In order to estimate this contribution, the same procedure described above is applied to simulated events of $\cPZ/\gamma^{\ast} \rightarrow \Pgt \Pgt$,
single top quark production, diboson production, and the part of the SM \ttbar background not included in the misidentified $\Pgt$ background estimate.

In order to estimate the misidentification  probability,
the hadronic multijet events are selected from a sample
triggered by at least one jet with $\pt >30$\GeVc, by requiring events to have at least two jets with $\pt >20$\GeVc and $|\eta | < 2.4$.
The triggering jet  is removed from the misidentification rate calculation in order to avoid a trigger bias.
The $\PW+\geq 1$~jet events are selected by requiring only one isolated muon with $\pt > 20$\GeVc
and $|\eta | < 2.1$, and at least one jet with $\pt > 20$\GeVc and $|\eta | < 2.4$.
The probability $w(\text{jet} \rightarrow \tauh)$ is evaluated from all jets in a sample enriched in QCD multijet events ($w_\mathrm{QCD}$),
and all jets in another sample enriched in $\PW+\geq 1$~jet events ($w_{\PW+\text{jets}}$).
The probability that a jet is misidentified as a $\Pgt$ jet as a function of jet \pt, $\eta$ and $R_\text{jet}$
is compared between simulated events (Z2 tune~\cite{c:z2}) and data, and a good agreement is found~\cite{c:tau-11-001}.

Jets in QCD multijet events are mainly gluon jets ($\simeq$ 75\% obtained from simulation),
while the jets in $\PW+\geq 1$~jet events are predominantly quark jets ($\simeq$64\% obtained from simulation), where $w_\mathrm{QCD}<w_{\PW+\text{jets}}$.
Since the quark and gluon jet composition in $\ell+\MET + \geq 3$~jet events lies between two categories of events, QCD multijet and $\PW+\geq 1$~jet events,
the $N^\text{misid}$ value is under- (over-) estimated by applying the $w_\mathrm{QCD}$ ($w_{\PW+\text{jets}}$) probability.
Thus, the $N^\text{misid}$ and its systematic uncertainty are estimated as in the following:

\begin{equation} \label{eq:fakes1}
N^\text{misid} = \frac{\sum^{N}_{i}~ \sum ^{n}_{j} w_{\PW+\text{jets},\,i}^{j}~+~\sum^{N}_{i}~ \sum ^{n}_{j} w_{\mathrm{QCD},\,i}^{j}}{2}
\end{equation}

\begin{equation} \label{eq:fakes2}
 \Delta N^\text{misid} = \frac{\sum^{N}_{i}~ \sum ^{n}_{j} w_{\PW+\text{jets},\,i}^{j}~-~\sum^{N}_{i}~ \sum ^{n}_{j} w_{\mathrm{QCD},\,i}^{j}}{2}
\end{equation}

The contribution of $N_\text{other}$ described earlier is subtracted from Eq.(\ref{eq:fakes1}).
Finally, the efficiency $\varepsilon _\mathrm{OS}$ of the OS requirement
obtained from simulated events is applied to obtain the misidentified $\Pgt$ background
$N_\mathrm{OS}^\text{misid} = \varepsilon_\mathrm{OS} \times N^\text{misid}$.
The estimated efficiencies for the $\Pe\tauh$ and $\Pgm\tauh$ final states are $\varepsilon _\mathrm{OS}=0.72 \pm 0.09 (\text{stat.}) \pm 0.02 (\text{syst.})$
and $\varepsilon_\mathrm{OS}=0.69 \pm 0.07 (\text{stat.})  \pm 0.03 (\text{syst.})$, respectively,
where the statistical uncertainty comes from the limited number of simulated events, and
the systematic uncertainty is taken as half of the difference of the efficiency
estimated from W+jets and lepton+jet \ttbar simulated events.

Other backgrounds in this analysis are
$\cPZ/\gamma^{\ast} \rightarrow \Pgt \Pgt$, single top quark production, diboson production, and the part of
the SM \ttbar background not included in the misidentified $\Pgt$ background, and are estimated from simulation.
Events from $\cPZ\rightarrow \Pe\Pe,\Pgm\Pgm$
are also taken into account because they
contain misidentified $\Pgt$ jets, where the misidentified
$\Pgt$ lepton can originate from an electron or muon misidentified as a $\Pgt$ jet.
The statistical uncertainties are due to the limited number of simulated events.

\section{Systematic uncertainties}
\label{sec:systematics}

Different sources of systematic uncertainties on the measurement of the cross section
due to signal selection efficiencies and backgrounds are considered, as shown in Table~\ref{tab:SummarySystematics}.
The main sources of systematic uncertainties are due to $\Pgt$ identification, \cPqb-tagging and mistagging efficiencies, jet energy scale (JES),
jet energy resolution (JER), \MET scale, and to the estimate of the misreconstructed $\Pgt$ background (from data).
The systematic uncertainties for the determination of the misidentified $\Pgt$ background are discussed in detail in Section~\ref{sec:background}.

The uncertainty on the $\Pgt$ jet identification includes contributions from
$\Pgt$ identification efficiency and $\ell~\rightarrow~\tauh$ ($\ell=\Pe,\Pgm$) misidentification.
The uncertainty on $\Pgt$ identification efficiency is estimated to be 6\% (from an updated measurement with respect to~\cite{c:tau-11-001}),
and it includes the uncertainty on charge determination which is estimated to be smaller than 1\%.
The uncertainty on the $\ell~\rightarrow~\tauh$
misidentification rate is estimated as the difference of $\Pgt$ misidentification rate measured in data and in simulated events, and is taken to be 15\%~\cite{c:tau-11-001}.
These uncertainties are applied to the simulated $\cPZ\rightarrow \Pe\Pe,\Pgm\Pgm$, and \ttbar dilepton background events.

The uncertainties related to \cPqb-tagging and mistagging efficiencies are estimated from
a variety of control samples enriched in \cPqb\ quarks, and the
data-to-simulation scale factors amount to $0.95\pm 0.06$ and $1.11\pm 0.11$, respectively~\cite{c:btv-11-001}.

The uncertainties on JES, JER, and \MET scale are estimated according to the prescription described in Ref.~\cite{jme-10-011}.
These uncertainties also take into account the uncertainty due to
the JES dependence on the parton flavor.
The uncertainty on JES is evaluated as a function of jet \pt and jet $\eta$.
The JES and JER uncertainties are propagated in order to estimate the uncertainty of the \MET scale.
An additional 10\% uncertainty on the contribution to \MET coming from the energy of particles that are not clustered into jets is also taken into account.

The theoretical uncertainty on the signal acceptance is estimated to be 4\%~\cite{c:top-11-002}.
It accounts for variations in the renormalization and factorization scales (2\%),
$\Pgt$ lepton and hadron decay modelling (2\%), top quark mass (1.6\%),
leptonic branching fractions of the \PW\ boson (1.7\%), and jet and \MET modelling (1\%).
Uncertainties on the PDFs are found to be negligible.

The uncertainty on the integrated luminosity is estimated to be 2.2\%~\cite{smp-12-008}.
The number of interactions per bunch crossing in the data (pile-up) is estimated
from the measured luminosity
in each bunch crossing
times an average total inelastic cross section (with an uncertainty of 6.5\%). The estimated number of interactions has a total uncertainty of approximately 8\%,
which corresponds to an overall uncertainty of the pile-up distribution.
The mean of pile-up in the data sample is about 5--6 interactions, with the uncertainty estimated conservatively by shifting the overall mean up or down by  0.6 interactions.

The lepton trigger, identification, and isolation efficiencies are measured
with the ``tag-and-probe" method
in events containing a lepton pair of
invariant mass between 76 and 106\GeVcc.
Within the precision of the present measurement, the scale factors
between efficiencies measured in data and in simulation are estimated to be equal to one.
The combined uncertainty on the electron (muon) trigger, identification and isolation efficiencies is 3\% (2\%).

Theoretical uncertainties on the cross sections of single top quark, diboson, and DY processes
are estimated as in Ref.~\cite{c:top-10-002}.
The uncertainties include the scale and PDF
uncertainties on theoretical cross sections.

\begin{table*}[htpb]
\caption{List of systematic uncertainties (in \%) on the cross section measurement.
The Best Linear Unbiased Estimation method~\cite{blue}
is used to combine the cross section measurements in the $\Pe\tauh$ and $\Pgm\tauh$ channels, with the corresponding weights.
Systematic uncertainties common to the two channels are assumed to be 100\% correlated.}
\label{tab:SummarySystematics}
\small
\setlength{\extrarowheight}{1.5pt}
\begin{center}
\begin{tabular}{|c|c|c|c|}
\hline
\hline
\multicolumn{1}{|c|}{Source} & \multicolumn{2}{c|}{Uncertainty [\%]} & \multicolumn{1}{c|}{ Combination [\%]}\\ \hline
                              &  $\Pe\tauh$        & $\Pgm \tauh$     &     \\
\hline
    $\tau$ misidentification background                               &       12.6        &       9.8      &  10.8   \\
\hline
    $\tau$ jet identification                            &       6.4         &       6.3      &  6.3 \\
\hline
    \cPqb-jet tagging, misidentification         &       5.3         &       5.3      &  5.3  \\
\hline
    jet energy scale, jet energy resolution, \MET        &       5.1         &       6.2      &  5.8  \\
\hline
    theoretical uncertainty on signal efficiency         &       4.0         &       4.0      &  4.0 \\
\hline
    pile-up modelling                                      &       2.3         &       2.3      &  2.3\\
\hline
    electron selection                                   &       3.1         &         0      &  1.1 \\
\hline
    muon selection                                       &       0           &       2.0      &  1.3 \\
\hline
    cross section of MC backgrounds                      &       1.6         &       1.4      &  1.5 \\
\hline
    luminosity                                           &       2.2         &       2.2      &  2.2 \\
\hline
\hline
    weight                                               &       0.38        &       0.62    & \multicolumn{1}{c|}{ $\chi^{2}/N_\mathrm{dof}=2.381/1$ }   \\
    &&&\multicolumn{1}{c|}{(p-value = 0.198) }\\
\hline
\hline
\end{tabular}

\end{center}
\end{table*}

\section{Cross section measurement}
\label{sec:xsec}

The number of events expected from the backgrounds, the number of signal events from \ttbar, and the number of
observed events after all selection cuts are summarized in Table~\ref{tab:SummaryEventYieldTauHadHPS}.
The statistical and systematic uncertainties
are also shown.

\begin{table*}[htpb]
\begin{center}
\caption{
Number of expected events for signal and backgrounds.
The background from ``misidentified $\Pgt$" is estimated from data, while the other backgrounds are estimated from simulation.
Statistical and systematic uncertainties are shown.
}
\label{tab:SummaryEventYieldTauHadHPS}
\setlength{\extrarowheight}{1.5pt}
\begin{tabular}{|c|c|c|}
\hline\hline
\multicolumn{1}{|c|}{Source} & \multicolumn{2}{c|}{N$_\text{events}$ ($\pm$ stat. $\pm$ syst.)} \\ \cline{2-3}
                             &         $\Pe\tauh$  &            $\Pgm\tauh$                          \\
\hline
$\ttbar \rightarrow {\PW\cPqb\PW\cPqb} \rightarrow \ell \Pgn \cPqb ~ \Pgt \Pgn \cPqb$ & 99.9 $\pm$ 3.0 $\pm$ 10.1 &   162.0 $\pm$ 4.0 $\pm$ 16.7  \\
misidentified $\tau$                 & 54.3 $\pm$ 6.4 $\pm$ 8.1 &    88.5 $\pm$ 8.9 $\pm$ 10.8   \\
$\cPZ/\gamma ^{\ast}\rightarrow \Pgt\Pgt$                          & 16.6  $\pm$ 3.3 $\pm$ 2.9 &   25.8  $\pm$ 4.3  $\pm$ 6.1  \\
$\ttbar \rightarrow {\PW\cPqb\PW\cPqb} \rightarrow \ell \Pgn \cPqb ~ \ell \Pgn \cPqb$ & 9.0  $\pm$ 0.9 $\pm$ 1.7 &   13.3  $\pm$ 1.2  $\pm$ 2.5  \\
$\cPZ/\gamma ^{\ast}\rightarrow {\Pe\Pe},\Pgm\Pgm$                         & 4.8  $\pm$ 1.8 $\pm$ 1.3 &   0.7  $\pm$ 0.7  $\pm$ 0.7   \\
Single top                                                      & 7.9  $\pm$ 0.4 $\pm$ 1.1 &   13.5  $\pm$ 0.5  $\pm$ 1.9  \\
VV                                                              & 1.3  $\pm$ 0.1 $\pm$ 0.2 &   2.0  $\pm$ 0.2  $\pm$ 0.3   \\
\hline
Total expected                                      &  193.9 $\pm$ 4.9 $\pm$ 18.0 &   306.1 $\pm$ 6.1 $\pm$ 27.9  \\
\hline
Data                                                            &  176 &   288         \\
\hline\hline
\end{tabular}
\end{center}
\end{table*}

The \ttbar production cross section measured from tau dilepton events is:

\begin{equation}
\label{eq:crosssection}
\sigma_{\ttbar} = \frac{N - B}{L \cdot A_\text{tot}},
\end{equation}

where $N$ is the number of observed candidate events, $B$ is the
estimate of the background, $L$ is the integrated luminosity.
The total acceptance $A_\text{tot}$ is
the product of all branching fractions, geometrical and kinematical acceptance, efficiencies for trigger,
lepton identification and the overall reconstruction efficiency, and it is evaluated with respect to the inclusive \ttbar sample.
After the OS requirement:
\begin{eqnarray}
\label{eq:efficiency}
A_\text{tot} (\Pe\tauh) = [0.0304 \pm 0.0009 (\text{stat.}) \pm 0.0031 (\text{syst.})]\%; \\
A_\text{tot} (\Pgm\tauh) = [0.0443 \pm 0.0011 (\text{stat.}) \pm 0.0047 (\text{syst.})]\%.
\end{eqnarray}

The statistical uncertainties are due to the limited number of simulated events and the systematic uncertainties are estimated by varying
all sources of systematics in Table~\ref{tab:SummarySystematics} affecting the signal (i.e., all uncertainties except for the luminosity and for the background).
All systematic and statistical uncertainties in Table~\ref{tab:SummaryEventYieldTauHadHPS} are propagated from Eq.(\ref{eq:crosssection})
to the final cross section measurement. The measured \ttbar cross section is:

\begin{eqnarray}
\label{eq:xsresult}
\sigma_{\ttbar} (\Pe\tauh) = 136 \pm 23 (\text{stat.}) \pm 23 (\text{syst.}) \pm 3 (\text{lumi.})\unit{pb}; \\
\sigma_{\ttbar} (\Pgm\tauh) = 147 \pm 18 (\text{stat.}) \pm 22 (\text{syst.}) \pm 3 (\text{lumi.})\unit{pb}.
\end{eqnarray}

The Best Linear Unbiased Estimation method~\cite{blue} is used to combine the cross section measurements
in the $\Pe\tauh$ and $\Pgm\tauh$ channels with the associated uncertainties and correlation factors.
Systematic uncertainties common to the two channels are assumed to be 100\% correlated.
The combined result is

\begin{eqnarray}
\label{eq:xsresultBLUE}
\sigma_{\ttbar} = 143 \pm 14 (\text{stat.}) \pm 22 (\text{syst.}) \pm 3 (\text{lumi.})\unit{pb},
\end{eqnarray}

in agreement with the measured values in the dilepton~\cite{c:top-11-002}
and lepton+jet~\cite{c:top-10-002,c:top-10-003} final states, and with the SM expectations
in the approximate NNLO calculation of
$163^{+7}_{-5} (\text{scale}) \pm 9 (\mathrm{PDF})$\unit{pb}~\cite{c:Kidonakis2}.

\section{Summary}
\label{sec:summ}

We present the first measurement of the \ttbar production cross section in the tau dilepton channel
$\ttbar\rightarrow (\ell\Pgn_\ell) (\tauh \Pgngt) \bbbar$, ($\ell=\Pe, \Pgm$)
with data samples corresponding to an integrated luminosity of 2.0--2.2\fbinv collected in proton-proton collisions at $\sqrt{s} =7$\TeV.
Events are selected by requiring the presence of one electron or muon, two or more jets (at least one jet is \cPqb\ tagged),
missing transverse energy, and one hadronically decaying $\Pgt$ lepton.
The largest background contributions come from events where one \PW\ boson is produced in association with jets,
and from $\ttbar \rightarrow \PWp\cPqb\PWm\cPaqb\rightarrow \ell\Pgn \cPqb ~\Pq\Paq^\prime\cPaqb$ events, where
one jet is misidentified as the $\Pgt$, and from $\cPZ\rightarrow\Pgt\Pgt$ events.
The measured
cross section
is $\sigma_{\ttbar} = 143 \pm 14 (\text{stat.}) \pm 22 (\text{syst.}) \pm 3 (\text{lumi.})\unit{pb}$,
in agreement with SM expectations.

\section*{Acknowledgments}
We wish to congratulate our colleagues in the CERN accelerator departments for the excellent performance of the LHC machine.
We thank the technical and administrative staff at CERN and other CMS institutes, and acknowledge support from: FMSR (Austria);
FNRS and FWO (Belgium); CNPq, CAPES, FAPERJ, and FAPESP (Brazil); MES (Bulgaria); CERN; CAS, MoST, and NSFC (China);
COLCIENCIAS (Colombia); MSES (Croatia); RPF (Cyprus); MoER, SF0690030s09 and ERDF (Estonia); Academy of Finland, MEC,
and HIP (Finland); CEA and CNRS/IN2P3 (France); BMBF, DFG, and HGF (Germany); GSRT (Greece); OTKA and NKTH (Hungary);
DAE and DST (India); IPM (Iran); SFI (Ireland); INFN (Italy); NRF and WCU (Korea); LAS (Lithuania); CINVESTAV, CONACYT, SEP,
and UASLP-FAI (Mexico); MSI (New Zealand); PAEC (Pakistan); MSHE and NSC (Poland); FCT (Portugal); JINR (Armenia, Belarus,
Georgia, Ukraine, Uzbekistan); MON, RosAtom, RAS and RFBR (Russia); MSTD (Serbia); MICINN and CPAN (Spain);
Swiss Funding Agencies (Switzerland); NSC (Taipei); TUBITAK and TAEK (Turkey); STFC (United Kingdom); DOE and NSF (USA).
Individuals have received support from the Marie-Curie programme and the European Research Council (European Union);
the Leventis Foundation; the A. P. Sloan Foundation; the Alexander von Humboldt Foundation; the Belgian Federal Science Policy Office;
the Fonds pour la Formation \`a la Recherche dans l'Industrie et dans l'Agriculture (FRIA-Belgium); the Agentschap voor Innovatie door
Wetenschap en Technologie (IWT-Belgium); the Council of Science and Industrial Research, India; and the HOMING PLUS programme
of Foundation for Polish Science, cofinanced from European Union, Regional Development Fund.

\bibliography{auto_generated}
\cleardoublepage \appendix\section{The CMS Collaboration \label{app:collab}}\begin{sloppypar}\hyphenpenalty=5000\widowpenalty=500\clubpenalty=5000\textbf{Yerevan Physics Institute,  Yerevan,  Armenia}\\*[0pt]
S.~Chatrchyan, V.~Khachatryan, A.M.~Sirunyan, A.~Tumasyan
\vskip\cmsinstskip
\textbf{Institut f\"{u}r Hochenergiephysik der OeAW,  Wien,  Austria}\\*[0pt]
W.~Adam, T.~Bergauer, M.~Dragicevic, J.~Er\"{o}, C.~Fabjan, M.~Friedl, R.~Fr\"{u}hwirth, V.M.~Ghete, J.~Hammer\cmsAuthorMark{1}, N.~H\"{o}rmann, J.~Hrubec, M.~Jeitler, W.~Kiesenhofer, M.~Krammer, D.~Liko, I.~Mikulec, M.~Pernicka$^{\textrm{\dag}}$, B.~Rahbaran, C.~Rohringer, H.~Rohringer, R.~Sch\"{o}fbeck, J.~Strauss, A.~Taurok, F.~Teischinger, P.~Wagner, W.~Waltenberger, G.~Walzel, E.~Widl, C.-E.~Wulz
% ...
\vskip\cmsinstskip
\dag:~Deceased\\
1:~~Also at CERN, European Organization for Nuclear Research, Geneva, Switzerland\\
% ...

\end{sloppypar}
\end{document}